\documentclass[twocolumn,amsmath,floatfix,prl,aps]{revtex4-1}

\usepackage{color}
\usepackage{amsmath}
\usepackage{pifont}   % Ding symbolsPHYSICAL REVIE% B UOLUME 16, NUMBER 4 15 AUGUST 1977
\usepackage{graphicx} % Include figure files
\usepackage{dcolumn}  % Align table columns on decimal point
\usepackage{bm}       % bold math
\usepackage{amsfonts} % Some more fonts
\usepackage{amssymb}  % More symbols
\usepackage{multirow} % Table functions
\usepackage{siunitx}
\usepackage[dvipsnames]{xcolor}
\usepackage[english]{babel}
\usepackage[autostyle, english = american]{csquotes}
\MakeOuterQuote{"}

\usepackage{placeins}

\usepackage{braket} % as suggested by marmot and this example answer https://tex.stackexchange.com/questions/316836/the-bra-ket-notation-in-latex

    \renewcommand{\v}[1]{\bm{\mathrm{#1}}}

\begin{document}
%\linenumbers

\title{Kohn-Sham-Proca equations for ultrafast exciton dynamics}
\author{J.~K. Dewhurst}
\affiliation{Max-Planck-Institut fur Mikrostrukturphysik Weinberg 2, D-06120 Halle, Germany}
\author{D. Gill}
\affiliation{Max-Born-Institute for Non-linear Optics and Short Pulse Spectroscopy, Max-Born Strasse 2A, 12489 Berlin, Germany}
\author{S. Shallcross}
\affiliation{Max-Born-Institute for Non-linear Optics and Short Pulse Spectroscopy, Max-Born Strasse 2A, 12489 Berlin, Germany}
\author{S. Sharma}
\email{sharma@mbi-berlin.de}
\affiliation{Max-Born-Institute for Non-linear Optics and Short Pulse Spectroscopy, Max-Born Strasse 2A, 12489 Berlin, Germany \\
Institute for theoretical solid-state physics, Freie Universit\"at Berlin, Arnimallee 14, 14195 Berlin, Germany}

\date{\today}

\begin{abstract}
A long-standing problem in time-dependent density functional theory has
been the absence of a functional able to capture excitonic physics under laser
pump conditions. Here we introduce a scheme of coupled Kohn-Sham and Proca
equations in a pump-probe setup that we show (i) produces linear-response
excitonic effects in the weak pump regime in excellent agreement with experiment,
but also (ii) captures excitonic physics in the highly non-linear regime of
ultrafast strong laser pumping. In particular "bleaching" (i.e. reduction) of
the excitonic weight and the appearance of excitonic side bands is demonstrated.
The approach is a procedural functional -- the Kohn-Sham and Proca equations are
simultaneously time-propagated -- allowing the straight-forward inclusion of,
for example, lattice and spin degrees of freedom into excitonic physics.
The functional is shown to have universal applicability to a wide range of
materials, and we also establish a relation between the parameters used in the
functional and the exciton Bohr radii of the materials.
\end{abstract}

\maketitle

%%%%%%%%%%%%%%%%%%%%%%%%%%%%%%%%%%%%%%%%%%%%%%%%%%%%%%%%%%%%%%%%%%%%%
%% Start the main part of the manuscript here.
%%%%%%%%%%%%%%%%%%%%%%%%%%%%%%%%%%%%%%%%%%%%%%%%%%%%%%%%%%%%%%%%%%

\section{Introduction}
Spin-tronics, magnonics, valley-tronics and excitonics~\cite{vzutic2004,schaibley2016}
are all potential alternatives to conventional charge-based electronics, which
are reaching their limits both in component and power densities.
In all of these fields, laser light is used to modulate and control the
fundamental excitations of a material to generate spin-, valley- or
exciton-current~\cite{melnikov2011,chen2018}. With a view towards the next
generation of optoelectronic devices, the field of transient
opto-{\it excitonics}, the ultrafast laser induced creation, dynamics, and
control of exciton dynamics, is one of the most
promising~\cite{sie2017,ouyang2020,kobayashi2023}.

A fully {\it ab-initio} method to describe such physics of light-matter
interaction is time-dependent density functional theory (TD-DFT)\cite{runge1984}:
it has already proved its strength in
the field of spin- and valley-tronics, predicting new phenomena as well as
providing a full microscopic understanding of the
physics \cite{krieger2015,dewhurst2018laser,Siegrist2019,
sharma2023thz,sharma2022valley,hofherr2020ultrafast,
schultze2014,Bertoni2016}. So far, however, the approximations used in TD-DFT are
unable to describe the time-dependent spectra of laser pumped excitons in solids.
In contrast for the linear-response regime (i.e., when the light perturbation
is very weak), TD-DFT has proved very successful in computing the excitonic
spectra of diverse materials \cite{sottile2003,reining2002,botti2004,sharma2011}. 
Following this approach, TD-DFT can be re-written as a Dyson equation linking the
Kohn-Sham non-interacting particle response to the full material response via
the exchange-correlation (XC) kernel,
$f_{\rm xc}({\bf r},{\bf r}',t-t')
 \equiv\delta v_{\rm xc}({\bf r}t)/\delta\rho({\bf r}'t')$,
where $v_{\rm xc}$ is the exchange-correlation potential and $\rho$ the
electronic charge density. To describe excitons in solids, this XC kernel in
the $q\rightarrow 0$ limit (i.e., for long wavelength pulses) must go as
$f_{\rm xc}\rightarrow 1/q^2$ \cite{onida2002,ghosez1997}. On this basis several
approximate XC kernels have been designed that successfully describe the
excitonic response of real materials \cite{sottile2003,reining2002,botti2004,sharma2011}.
However, since these approximations are exclusively in the form of XC kernels,
their use is restricted to linear regime and cannot be extended to strong field
pumped time-dependent excitons. It is thus, at present, not possible to have a
unified method that can treat linear and strong-field excitons and most
importantly, it is not possible to study excitronics (the dynamics of pumped
excitons) using the most prominent method of choice, TD-DFT.

In this work, we describe a real-time TD-DFT method for solids that captures the
dynamics of excitons by coupling the electronic Kohn-Sham (KS) equations to an
effective vector potential ${\bf A}_{\rm xc}$, time-dependent but not spatially
varying. We show that  ${\bf A}_{\rm xc}$ as a solution of a {\em Proca equation},
which is the Maxwell equation containing a mass term, generates the functional
for ${\bf A}_{\rm xc}$ that correctly describes the excitonic response in the
linear as well as strong field regime. We
demonstrate this by theoretical arguments as well as by numerical results; by
comparison with experiment we demonstrate that in the weak pump-probe regime, the
Kohn-Sham-Proca (KSP) functional correctly gives the linear-response limit and also good
agreement with experiments for the dynamics of strongly pumped exciton.

\section{Method}
Real-time TD-DFT \cite{runge1984,sharma2014} rigorously maps the computationally
intractable problem of interacting electrons to a KS system of non-interacting
electrons in an effective potential. The time-dependent KS equation is:
\begin{align}  
 i\frac{\partial \psi_{j}({\bf r},t)}{\partial t}=
 \Bigg[\frac{1}{2}\Big(-i{\nabla}-\frac{1}{c}\big({\bf A}(t)
 +{\bf A}_{\rm xc}(t)\big)\Big)^2+v_{s}({\bf r},t) \Bigg]
 \psi_{j}({\bf r},t),
 \label{e:TDKS}
\end{align}
where $\psi_j$ is a KS spinor orbital and the effective KS potential
$v_{s}({\bf r},t) \equiv v({\bf r},t)+v_{\rm H}({\bf r},t)+v_{\rm xc}({\bf r},t)$
consists of the external potential $v$, the classical electrostatic Hartree
potential $v_{\rm H}$ and the exchange-correlation (XC) potential $v_{\rm xc}$.
The vector potential ${\bf A}(t)$ represents the applied laser field within the
dipole approximation (i.e., the spatial dependence of the vector potential is
absent) and ${\bf A}_{\rm xc}(t)$  the XC vector potential.
Note that, unless otherwise stated, atomic units are used throughout.

Within TD-DFT two conceptual routes exist to calculate the response in solids:
(a) linear-response formalism and (b) real-time TD-DFT. The former case is valid
for very weak perturbation where the TD-DFT equation can be cast into a Dyson-like
equation to determine the response,
requiring $f_{\rm xc} \to 1/q^2$ as $q \rightarrow 0$ for the
XC kernel \cite{sharma2011,sharma2014,reining2002,sottile2003}.
It is important to note that in this case the XC kernel is $a_2/q^2$ where
the scaling $a_2 < 1$ for weak and $a_2> 1$ for strongly bound excitons
(see Ref. 14 in Ref. \cite{sharma2011}).

Looking at this one would imagine that since
$f_{\rm xc}({\bf r},{\bf r}',t-t')\equiv
 \delta v_{\rm xc}({\bf r}t)/\delta\rho({\bf r}'t')$,
one could choose a corresponding form for $v_{\rm xc}$ to describe excitons in the
latter form of TD-DFT, i.e. real-time TD-DFT. However, this assumption is
invalid at a practical level for periodic solids -- excitons cannot be
treated by merely improving the lattice-periodic $v_{\rm xc}$ since the
$q \rightarrow 0$ limit has already been taken. Any part of the response which
is finite in this limit is necessarily excluded.

This problem can however, be solved in the following manner: the crucial
quantity appearing in the response formalism is $V({\bf q})\chi_s({\bf q})$,
where $\chi_s({\bf q})$ is the KS response function. In the $q\to 0$ limit,
this becomes an indeterminate form and can be replaced by an expectation value
of the momentum operator $\hat{\bf p}=-i\nabla$.
Emulating this with a lattice-periodic calculation thus requires coupling to the
variable conjugate to $\hat{\bf p}$, namely the XC vector potential
${\bf A}_{\rm xc}(t)$ \cite{vignale1988,pina2005,sun2021}. It follows that
within TD-DFT, excitons and their dynamics can be described by the time
dependence of the total current. The procedure for determining the
optical response then is to solve the time-dependent Hamiltonian with
${\bf A}_{\rm xc}(t)$ and, using the KS orbitals obtained as a solution to this
Hamiltonian, evaluating the the total current $\v J(t)$, which
itself is obtained by integrating the current density $\v j(\v r,t)$ in the
unit cell\cite{yabana12}. This gauge-invariant current density is given by:
\begin{align}
 {\bf j}({\bf r},t)=
 {\rm Im}\sum_j^{\rm occ}\psi_j^{\dag}({\bf r},t){\nabla}\psi_j({\bf r},t)
 -\frac{1}{c}\big({\bf A}(t)+{\bf A}_{\rm xc}(t)\big)\rho({\bf r},t).
\label{e:j}
\end{align}
The Fourier transform of this current to frequency space is used to generate
the optical conductivity, $\sigma(\omega)$, via the linear-response equation
${\bf J(\omega)}=\sigma(\omega) {\bf E(\omega)}$, where
${\bf E}(t)=-(1/c)\,\partial {\bf A}(t)/\partial t$ is the physical
(or external) electric field. This conductivity is then used to obtain
the dielectric function \cite{yabana2006},
$\varepsilon(\omega) = 1+4\pi i \sigma(\omega)/\omega$.
Since the full current is used to obtain the response,
local field effects are automatically included\cite{yabana12}.

One still requires an explicit functional form for ${\bf A}_{\rm xc}$ capable of
capturing the excitonic response of real solids. All approximate exchange-correlation
functional forms are ultimately choices, often based on heuristic arguments.
In our case, the physical vector potential of electromagnetism obeys Maxwell's
equation, and so a natural guess might be to obtain  ${\bf A}_{\rm xc}$ as the
solution of a Maxwell-like equation. This was recently implemented and tested by
Sun, {\it et al.}\cite{sun2021}. This approach, however, results in
an uncontrolled divergence of the time-dependent current.

%which was mitigated by an unphysical
%scaling \cite{my-book} of the XC potential, $v_{\rm xc}$, and an
%artificial time truncation of the dynamics.
%
%At this point it is crucial to ask an essential question in general and in
%particular about the work of Sun{\it et al.};

Why does the coupled TD-DFT Maxwell approach of Sun, {\it et al.} diverge? The
reason for this is not any essential divergence of coupled Kohn-Sham-Maxwell
equations approach: this method has been used successfully before for finite
systems\cite{yabana}. In this case the electromagnetic field radiates energy
away from the center and consequently acts as a dissipative system\cite{yabana}.
However, in the case of periodic boundary conditions with a spatially
independent vector potential, all unit-cells and hence the entire system behave
coherently. The vector potential ${\bf A}(t)$ feeds back into the TD-DFT
equations to increase the current, which itself enhances the vector potential
via the Maxwell equation, and so on, resulting in a numerically unstable procedure.

%This can be demonstrated numerically where a model Gaussian-shaped current (in Fig.~\ref{fig:fig1}(a)) results in a divergent ${\bf A}_{\rm xc}$ (red line Fig.~\ref{fig:fig1}(b)).  

\begin{figure}[ht]
 \centering
 \includegraphics[width=0.9\columnwidth,clip]{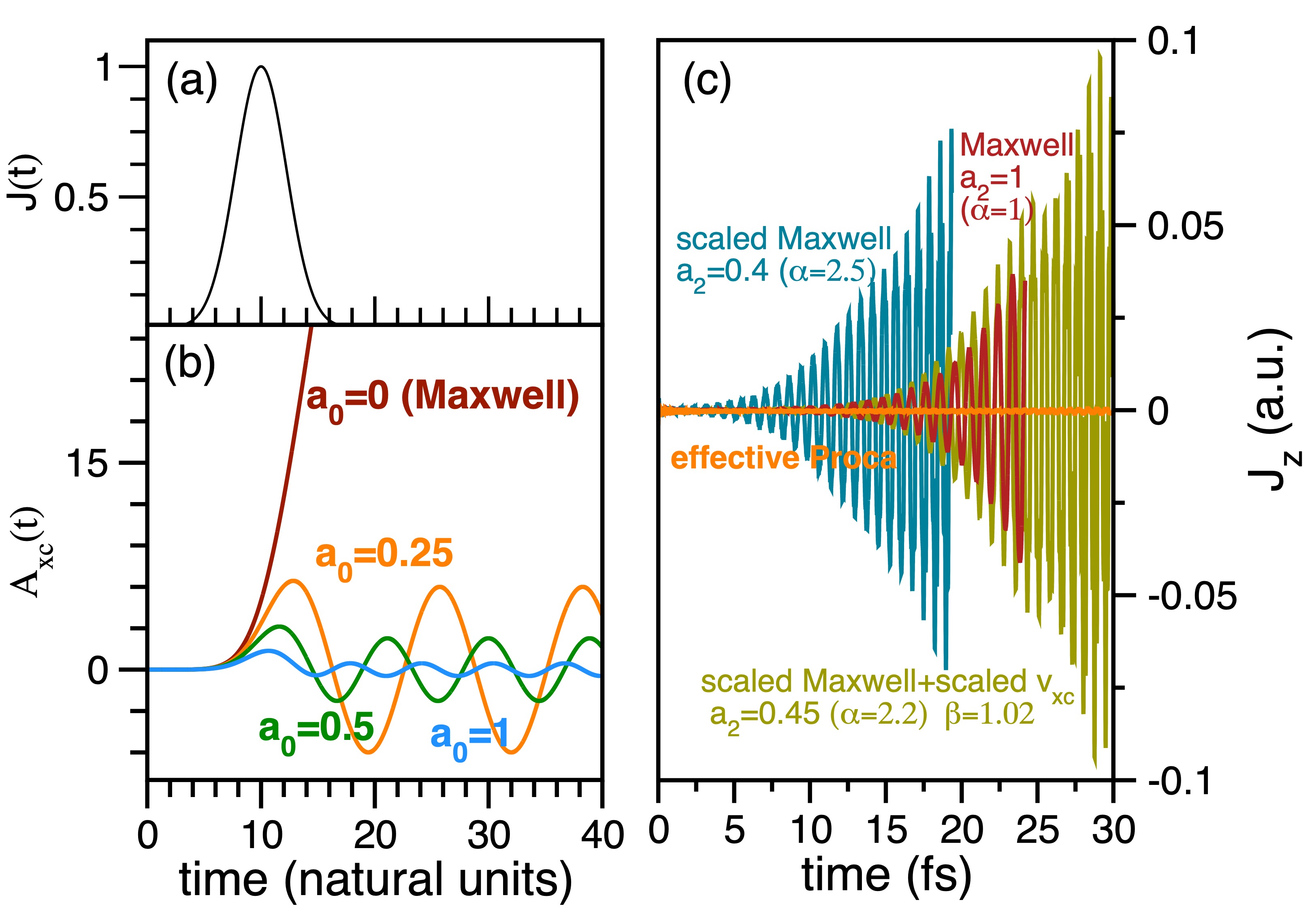}
 \caption{(a) current ${\bf J}(t)$. (b) ${\bf A}_{\rm xc}(t)$ obtained by solving
 the Proca (see Eq. (\ref{e:proca})) and Maxwell equations (obtained by setting
 $a_0=0$ and $a_2=1$ in Eq. (\ref{e:proca})) with ${\bf J}(t)$ from
 panel (a). (c) ${\bf A}_{\rm xc}(t)$ for LiF obtained by solving coupled
 Eqs. (\ref{e:TDKS}) and ~\ref{e:proca}. For all values of $a_2$, as well as
 for scaling of the XC potential $\beta v_{\rm xc}$ as suggested in
 Ref.~\cite{sun2021}, the current diverges when the Kohn-Sham system is
 coupled to an effective Maxwell equation. In contrast for the
 Kohn-Sham-Proca scheme ($a_0=0.25$) the current remains finite and
 oscillatory at all times.}
 \label{fig:fig1}
\end{figure}

This can be demonstrated; we first examine the solutions of the massless Maxwell equation
and as a first step look at the simplest case with no current term $J(t)$, i.e. the "free solution". 
This yields for the Maxwell equation a solution of the form $A_{\rm xc}(t)=c_0+c_1 t$ and for the
Proca equation $A_{\rm xc}(t)=c_1\sin\big(\sqrt{a_0/a_2}\,t+c_0\big)$. 
Strikingly, the solution of the Maxwell equation diverges while that of 
the Proca equation oscillates at frequency $\omega=\sqrt{a_0/a_2}$. As a second step one can 
then choose a simplistic form for $J(t)$, like a Gaussian (Fig.~\ref{fig:fig1}(a)), 
to study the solution for full non-homogeneous equation. 
This is shown in Fig.~\ref{fig:fig1}(b) from which  it is clear that the 
features of free solution are inherited by the full non-homogeneous equations; solution of 
Maxwell equaton diverges, while that of Proca equation oscillates with frequency proportional to the
ratio $\sqrt{a_0/a_2}$.

In a realistic case, this current ${\bf J}(t)$, rather than being fixed, arises
as the solution of Kohn-Sham equations coupled to, and simultaneously
propagated with the Maxwell or Proca equation. As
demonstrated in Fig.~\ref{fig:fig1}(c) for the case of LiF, the
current -- as well as ${\bf A}_{\rm xc}$ -- diverges for the Maxwell
equation, which cannot be cured by any scaling of the XC potential $v_{\rm xc}$, but can be 
delayed in time. This delay allows for a short window
in which a response can be determined and was done in the work of Sun {\it et al.}
However, this makes the procedure invalid for the study of real-time laser
pumped excitons. 
The Proca equation solution, on the other hand, remains finite and oscillatory. All
calculations in this work are performed using state-of-the-art full-potential
linearized augmented plane wave method \cite{singh} as implemented in the Elk
code \cite{elk,dewhurst2016} (for further details see the SI).

% what is alpha?

Can we nevertheless improve on this functional generated via the Maxwell
equation to make it both stable and have wide applicability?
For this we look at a more general form of the Maxwell equation itself, namely
the so-called Proca equation that includes mass term and which, in the absence of
spatial variations, is
\begin{align}
 a_2\frac{\partial^2}{\partial t^2}{\bf A}_{\rm xc}(t)+a_0{\bf A}_{\rm xc}(t)=
 \frac{4\pi c}{\Omega}{\bf J}(t),
\label{e:proca}
\end{align}
where $a_0$ and $a_2$ are parameters (the Maxwell equation is obtained by
setting $a_2=1$ and $a_0=0$), and $\Omega$ is the unit cell volume. Note the
prefactor of $c$ in Eq. (\ref{e:proca}): this along with the prefactor of
$1/c$ in Eq. (\ref{e:TDKS}) ensures that
the effects generated by ${\bf A}_{\rm xc}$ are {\em non-relativistic} in nature.
This is analogous to the XC magnetic field ${\bf B}_{\rm xc}$, arising from
interactions between electrons, rather than being relativistic in origin.

%From this equation one can show
%(see SI for full proof) that by appropriate choice of $a_0$ the functional
%can be made totally local i.e. like the local density approximation. This
%implies that by varying $a_0$ one goes from highly non-local ($1/q^2$) to a
%totally local functional. Thus one can adjust $a_0$ to form a hybrid
%functional that damps $1/q^2$ behavior to fully get rid of the divergence. 
%
Thus we propose a novel functional form for ${\bf A}_{\rm xc}$ by
coupling TD-DFT to an effective Proca equation. It should be stressed that
while the Proca equation entails that the XC field is massive this should not be
interpreted as a physical mass but rather an effective field, the aim of
which is to reproduce the time-dependent current.

This type of functional is {\em procedural} in that ${\bf A}_{\rm xc}$ is
determined not from a simple formula, but rather by the procedure of computing the
instantaneous ${\bf J}(t)$ from the KS wave-function at each time-step and using
this current to advance the Proca equation (\ref{e:proca}) to the next time-step.
Since the Proca equation is a second-order differential equation, two initial
values are required. As ${\bf A}_{\rm xc}$ describes the response of the
electron system to the external electromagnetic field it is natural to choose
${\bf A}_{\rm xc}(0)=\dot{\bf A}_{\rm xc}(0)=0$. If we convert
${\bf A}_{\rm xc}$ into its electric field form
${\bf E}_{\rm xc}(t)=-(1/c)\,\partial {\bf A}_{\rm xc}(t)/\partial t$,
we find this satisfies
\begin{align}\label{e:efield}
 a_2\frac{\partial}{\partial t}{\bf E}_{\rm xc}(t) %+a_1 {\bf E}_{\rm xc}(t)
 +a_0\int_0^t dt'\,{\bf E}_{\rm xc}(t')= -\frac{4 \pi}{\Omega} {\bf J}(t),
\end{align}
from which $v_{\rm xc}({\bf r},t)=-{\bf E}_{\rm xc}(t)\cdot {\bf r}$ would be the
corresponding scalar potential, which is clearly not lattice-periodic.
Once again, we can see from the lack of factors of $c$ this exchange-correlation
potential is non-relativistic in origin.
It is interesting to note that this a rare example of a functional specific to
TD-DFT, instead of a ground-state functional used adiabatically. In fact, it has
no effect on the ground-state because there the total current is zero. Furthermore,
it has memory effects because it does not depend exclusively upon the
instantaneous ${\bf J}(t)$, but also upon its own state.

It is not sufficient for an approximate functional to be stable. It must also
exhibit {\em universality}, in other words it should be predictive for a wide
range of materials, including hypothetical examples, while being free of adjustable
parameters. As we will see below, the Proca functional satisfies this condition
since the parameters $a_0$ and $a_2$ can be determined for any given material
merely from its band gap. Furthermore one can ascribe physical meaning to
the parameters used in the Proca equation.
To do this, let us suppose that spatial variations are also included in the
Proca equation. In the static limit, this becomes
$\nabla^2 v_{\rm xc}=(1/\lambda^2)v_{\rm xc}$, where $\lambda=\sqrt{a_2/a_0}$,
the solution of which is the Yukawa potential
$v_{\rm xc}(r)=(1/r)\exp(-r/\lambda)$.
Thus $\lambda$ provides a natural length-scale for the functional which,
as we will demonstrate, is proportional to the excitonic Bohr radius.
In the work of Sun, {\it et al.}\cite{sun2021}, $a_0$ is implicitly zero
and thus their solutions correspond to weakly bound excitons of infinite extent. 

It should be noted that this form of the Proca equation can be further generalized by adding a 
term proportional to the first derivative of the vector potential:
$a_1 {\partial {\bf A}_{\rm xc}(t)}/{\partial t}$. This acts as a dissipative term and would serve to 
decrease the life-time of the exciton, which would be useful in very long time simulations. As an ultimate
generalization, the parameters $a_0$, $a_1$, and $a_2$ could be also made $3 \times 3$ matrices accounting for 
directional inhomogeneity in materials.

%%%%%%%%%%%%%%%%%%%%%%%%%%%%%%%%%%%%%%%%%%%%%
% Results static
%%%%%%%%%%%%%%%%%%%%%%%%%%%%%%%%%%%%%%%%%%%%%

In the present work we simulate a realistic experimental pump-probe setup;
${\bf J(\omega)}$ is obtained by taking the difference between the currents
obtained from two calculations -- one with the pump pulse alone and one with
both pump and probe pulse. We associate the time delay between the pump and the
probe pulse with the time at which the material is probed after pumping.

\begin{figure}[t!]
 \centering
 \includegraphics[width=0.75\columnwidth,clip]{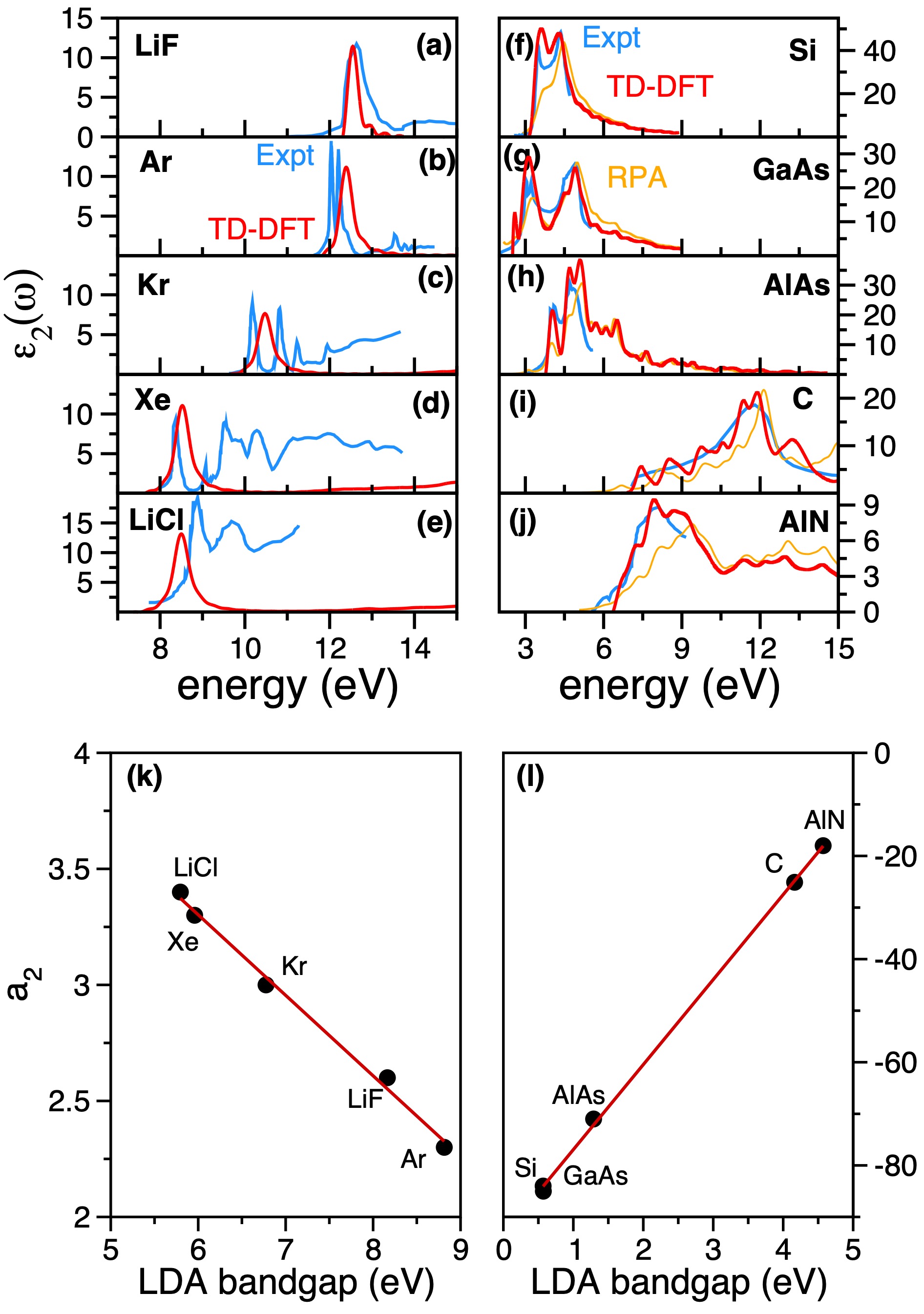}
 \caption{Imaginary part of the dielectric tensor, $\varepsilon_2(\omega)$, as
  a function of photon energy (in eV) for strongly bound (a-e) and weakly
  bound (f-j) excitonic materials. Experimental data are taken from the
  following sources: LiF~\cite{roessler1967}, Ar~\cite{saile1976},
  Kr~\cite{baldini1962}, Xe~\cite{saile1976}, LiCl~\cite{knox1959},
  Si~\cite{lautenschlager1987}, GaAs~\cite{lautenschlager1987GaAs},
  AlAs~\cite{garriga1993}, C~\cite{phillip1964}, AlN~\cite{cimalla2005}
  (see Table I of the SI for details) with random phase approximation (orange)
  results also shown also for comparison. Variation of the $a_2$ parameter as a
  function of Kohn-Sham band gap (in eV) (k) for strongly bound excitonic and
  (l) weakly bound excitonic materials. The red line is the fit to the data.}
  \label{fig:fig2}
\end{figure}

\begin{figure}[t!]
 \centering
 \includegraphics[width=0.8\columnwidth]{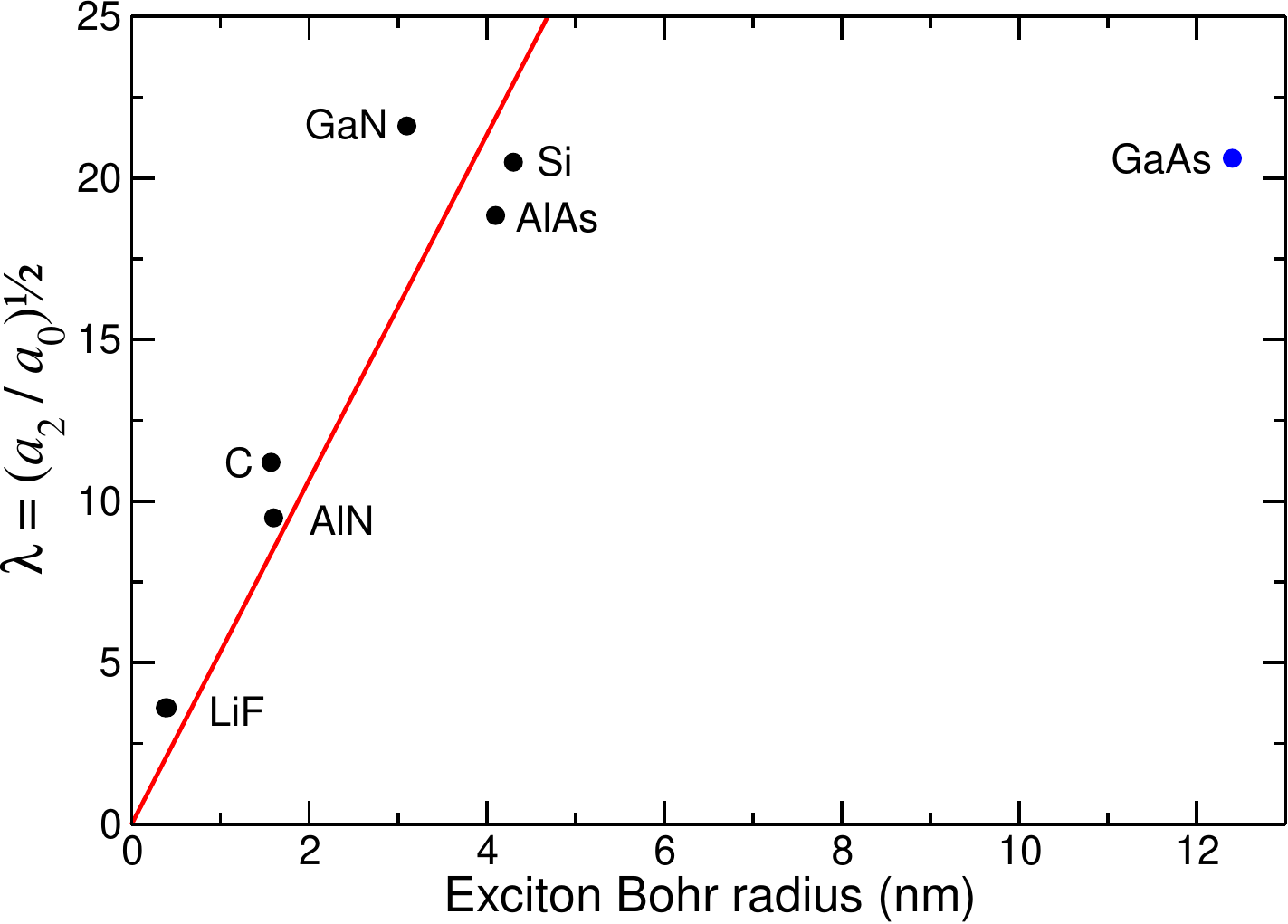}
 \caption{Plot of exciton Bohr radius vs. length-scale of the ${\bf A}_{\rm xc}$
 functional given by $\lambda=\sqrt{a_2/a_0}$. The radii are obtained from
 Ref. \cite{nano12142501} and references therein; and
 Refs. \cite{Puschnig2002,Shahrokhi2018,Okushi2005}. The line is a fit to the
 data (excluding the GaAs point) and passing through zero.}\label{fig:ex_rad}
\end{figure}

\section{Excitonic physics in weak pump pulse regime}
We now explore the performance of the coupled KSP equations as
applied to study materials under pump-probe conditions. We first study the case
of a weak pump pulse of duration of 12~fs with the material being probed 50~fs
after pumping. The pump and the probe pulses are kept to the same low fluence
of 0.001~mJ/cm$^2$.

Presented in Fig.~\ref{fig:fig2} for ten different materials, is the
absorption (i.e., the imaginary part of the dielectric function
$\varepsilon(\omega)$) calculated using the KSP pump-probe
scheme, using the random phase approximation (RPA) within linear-response
formalism of TD-DFT (i.e. by solving the Dyson equation), and experimental
linear-response absorption spectra\color{black}. The wide-gap insulators,
panels (a-e), all exhibit strongly bound exciton resonances with binding
energies ranging from 0.94~eV to 1.7~eV, and the KSP procedural functional is
seen to provide excellent agreement for the main excitonic peak position and
its height. For noble gas solids, panels (b-d), the KSP scheme well captures
the main exciton, but not the excited states of the exciton
(i.e. the Rydberg series). In the case of  LiCl, panel (e), calculations are
performed for an ideal crystal structure while experiments are performed for
thin films grown on LiF, leading to disagreement at above band-gap energies.
For the semi-conductors, panels (f-j), excitonic physics shows up instead as a
pronounced lowering and redistribution of the spectral weight as compared
to RPA, and here once again the KSP scheme is seen to yield excellent agreement
with the experimental linear-response spectra.
Overall, we thus conclude that the KSP procedural functional in a pump-probe
scheme captures very well the excitonic physics, correctly reproducing the
linear-response regime in the weak pumping limit.

\section{Universal form for the Proca equation}
In obtaining these results the
KSP procedure involved two material-dependent parameters, $a_0$ and $a_2$.
An indispensable feature of any XC approximation within DFT is its ability
to predict the properties of unknown materials or, known materials under novel
conditions, any free parameters must be internally determinable. As we now
show, the Proca parameters have precisely this feature. The term $a_0$ we found to
be material independent but material class dependent: it is fixed to $0.2$ for
the case of strongly bound excitons and $-0.2$ for the case of weakly bound
excitons. The results of our fit for $a_2$ are shown in
Fig.~\ref{fig:fig2}(k-l) revealing that this parameter depends linearly on
the band gap. This linear form represents a universal relation between a
calculable theoretical quantity -- the gap -- and the value of $a_2$ that
yields accurate treatment of the excitonic response, guaranteeing a robust
internally determinable scheme for the unknown parameters and hence a
{\it bona fide} universal procedural functional. This parameterization is
robust to the replacement of the LDA gap by its experimental value as shown
in the SI. More importantly, replacement of the LDA $v_{\rm xc}$ by that of a
different approximation, for example the generalized gradient approximation
(GGA), has negligible impact on our results (this is also shown in the SI),
reflecting the fact that excitonic physics in periodic solids is governed
by ${\bf A}_{\rm xc}$ and not $v_{\rm xc}$. We would note that caution must be
required for extended solids whose gap places the material intermediate
between strongly bound and weakly bound excitonic physics.

One can also compare the natural length-scale of the functional
$\lambda=\sqrt{a_2/a_0}$ with the average excitonic Bohr radius. This
comparison is plotted in Fig.~\ref{fig:ex_rad}. We find a linear relationship
between $\lambda$ and the exciton Bohr radius with the exception of
GaAs, which is an outlier. This is evidence that the parameters $a_0$ and
$a_2$ define both a current coupling strength as well as a
screening length.

%%%%%%%%%%%%%%%%%%%%%%%%%%%%%%%%%%%%%%%%%%%%%
% Results Dynamic
%%%%%%%%%%%%%%%%%%%%%%%%%%%%%%%%%%%%%%%%%%%%%

\begin{figure}[t!]
 \centering
 \includegraphics[width=\columnwidth,clip]{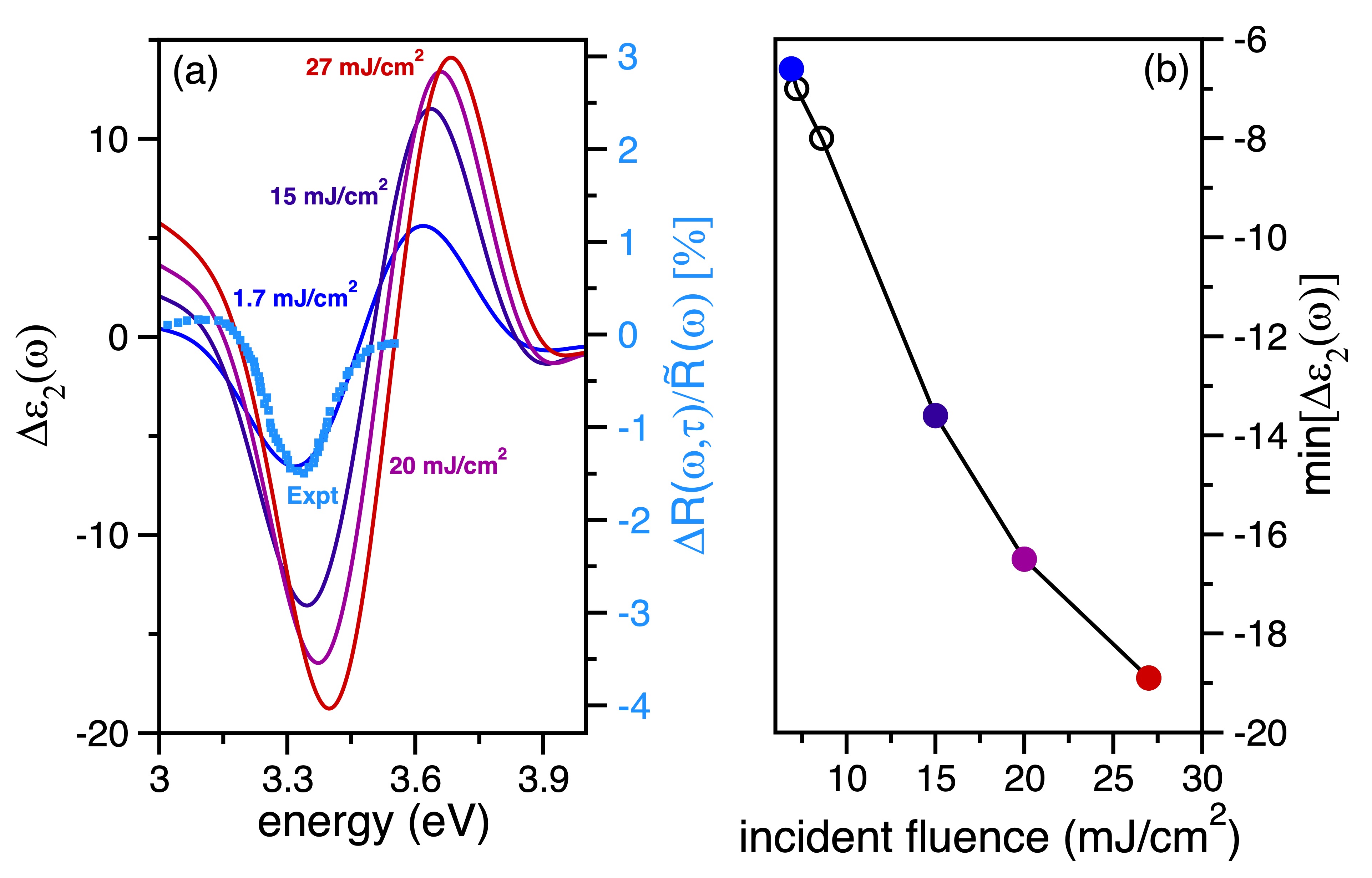}
 \caption{For bulk Si, (a) the variation of change in the imaginary part of the
  dielectric tensor due to laser pumping $\Delta \varepsilon_2(\omega)$
  as a function of photon energy (in eV), shown for different values of incident
  pump fluence (in mJ/cm$^2$) calculated using KSP procedural functional.
  (b) The minimum value of
  $\Delta \varepsilon_2(\omega)$ %($\min[\Delta \varepsilon_2(\omega)]$)
  as a function of incident fluence. Shown also are the experimental
  results\cite{sangalli2016} for percentage change in reflectance upon low
  fluence laser pumping.}
  \label{fig:fig3}
\end{figure}

\section{Excitons under strong pump laser conditions}
To study strong laser pumped excitons we proceed by pumping bulk Si with
short laser pulses (duration 12~fs) of varying fluence, 3 to 4 orders of
magnitude larger than the probe pulse fluence. The system is again probed
50~fs after pumping. The change in absorption for Si is shown in
Fig. \ref{fig:fig3}(a), revealing a bleaching effect (decrease in absorption)
between 3.3-3.6~eV and the appearance of side-bands (i.e., increased absorption)
above and below the bleached excitonic response. These features are in line
with experiments in Si~\cite{sangalli2016} -- data also shown in
Fig.~\ref{fig:fig3}(a) -- that report bleaching between 3.3-3.5~eV and the
appearance of side bands. Similar physics has also been reported in laser
pumped excitons in 2D materials~\cite{nie2014,mai2014,sie2017,ouyang2020,
kobayashi2023,smejkal2020,lucchini2021,sim2013}. While we cannot directly
compare the calculated amplitude with the experimental data, which is the
percentage change in reflectivity, if we track the minimum of the
$\Delta \varepsilon$ between 3-3.5~eV, see Fig.~\ref{fig:fig3}(b), we find a
linear behavior initially exactly as in experiment \cite{sangalli2016}.
This behavior deviates from linearity in very strong field regime. 

The exchange correlation vector potential evidently does not alter the
ground-state of the material, and hence the ground-state gap of the system is
not affected by this formalism. The transient optical gap, however, which will
not change in the limit of weak field pumping, does change upon strong field
pumping and contributes, together with transient changes in screening, towards
the shift of the excitonic peak in energy range 3.3-3.6~eV.
That two internally determinable parameters -- $a_0$ and $a_2$ -- can capture
both the excitons in the weak field regime as well as the physics of exciton
peak bleaching under strong field pumping represents a striking demonstration
of the power of the KSP procedural functional.

\section{Discussion}
We have presented a procedural functional KSP scheme that we have shown
(i) has all parameters internally determinable due to a universal relation
between the Proca mass and the band gap, (ii) possesses memory of the dynamics,
and (iii) captures key features of excitonic physics both in the weak as well
as in the strong laser pumping regimes. The calculated response under weak
laser pumping exhibits excellent agreement with experimental linear-response
spectra for a wide class of materials, while for Si we have shown the KSP
scheme captures key features of strong laser pumped excitons (bleaching and the
appearance of excitonic side bands).

In addition to this utility for the purely electronic system, the simultaneous
time propagation of Kohn-Sham and Proca equations renders it easy to include
additional degrees of freedom. Our approach thus both solves a crucial problem
of TD-DFT -- the ability to treat excitonic physics both in weak pumping as
well as in the highly non-equilibrium strong pumping time-dependent
case -- but also opens rich new possibilities for the coupling of excitons to
quasi-particles such as phonons and magnons, under laser pump conditions.

\section{Acknowledgements}
S. Sharma, JKD and DG would like to thank the DFG for funding through
project-ID 328545488 TRR227 (project A04). Sharma and Shallcross would like to
thank Leibniz Professorin Program (SAW P118/2021) for funding. 

%\bibliography{exciton}
%merlin.mbs apsrev4-1.bst 2010-07-25 4.21a (PWD, AO, DPC) hacked
%Control: key (0)
%Control: author (8) initials jnrlst
%Control: editor formatted (1) identically to author
%Control: production of article title (-1) disabled
%Control: page (0) single
%Control: year (1) truncated
%Control: production of eprint (0) enabled
%

\end{document}